\documentclass[11pt,a4paper]{article}
\usepackage[margin=2.4cm]{geometry}
\usepackage{amsmath,amssymb}
\usepackage{amsthm}
\usepackage{enumitem}
\usepackage{placeins}
\newtheorem{proposition}{Proposition}
\newcommand{\thisdoc}{paper}
\usepackage{booktabs}
\usepackage{array}
\usepackage{tikz}
\usepackage[round,authoryear]{natbib}
\usepackage[colorlinks=true,citecolor=blue!60!black,linkcolor=blue!60!black,urlcolor=blue!60!black]{hyperref}
\usepackage[expansion=false]{microtype}
\title{\vspace{-1.6cm}Nested logit: model structure and normalisation}
\author{Stephane Hess, University of Leeds}
\date{\today}

\newif\ifacademic
\academictrue

\begin{document}
\maketitle

\begin{abstract}
\noindent Nested logit remains a widely used model, especially in applied work. However, there remains substantial confusion surrounding the existence of different versions of the model, different normalisations for the same version, and different software implementations. This paper argues that the labels in common use, RU1 and RU2, and the name non-normalised nested logit (NNNL) itself, have contributed to the confusion. The model is written in two forms: one in which utilities are divided by the structural parameter of the nest that contains them, which is consistent with utility maximisation by construction, and one without that division, known as NNNL. The difference between these forms is not one of normalisation. Normalising at the top or at the bottom gives the same model, and NNNL is itself normalised. What distinguishes the forms is whether the division is performed. In short, normalisation changes the units, whereas division changes the model. The undivided form coincides with the utility-consistent one only under specific conditions, and otherwise is consistent with utility maximisation only under utilities other than those specified, in which generic coefficients become nest-specific. The paper highlights that RU1 has been used for three different specifications, compares how widely used software packages normalise and report their parameters, and illustrates the consequences with calculated examples. The recommendation is that models be described by what they do, rather than by these labels.
\end{abstract}




\ifacademic\section{Introduction and model context}\fi\label{sec:intro}

The Nested Logit (NL) model has a long tradition in choice modelling, and is a widely used structure especially in applied work. This \thisdoc{} does not discuss the history of the development of NL, which is explored in detail in \citet{ortuzar2001}, or revisit the motivation and derivation of the model, for which the reader is referred to textbook expositions in \citet[Ch.~10]{benakiva1985}, \citet[Ch.~4]{train2009}, \citet[Section~7.4]{ortuzar2011} and \citet[Section~4.6]{hensher2015}, or to the original work of \citet{williams1977}, \citet{dalyzachary1978} and \citet{mcfadden1978}. Rather, the focus is on normalisation and on model structure and implementation. The \thisdoc{} seeks to specifically address confusion surrounding the description of models as RU1 and RU2, and the relationship between these and a version described by \citet{koppelman1998a} as a non-normalised nested logit (NNNL) model.

\ifacademic
Despite a substantial literature on the topic, confusion remains, and the labels in common use contribute to it: RU1 has been read both as a normalisation and as a different model, and the name NNNL suggests that a normalisation is missing, although the model is normalised. This paper contributes to the literature by separating the two questions that the labels conflate, namely which scale is fixed and whether the utilities are divided by the structural parameter of the nest that contains them. Normalisation changes the units, whereas division changes the model. The central consequence is that coefficients specified as generic in the undivided form become, in its utility-consistent equivalent, specific to the nests, in proportion to their structural parameters. The model may then be consistent with utility maximisation, but not with the utilities the analyst specified (Proposition~\ref{prop:nnnl} in Section~\ref{sec:nnnl}), and its loss of translational invariance with respect to those utilities is the visible symptom. It also sets out what each label has been used for, and what widely used software actually estimates and reports. Finally, it shows, with calculated examples, when the undivided form reproduces the utility-consistent one and what happens when it does not. The paper concludes that the labels should be replaced by a statement of what a specific implementation of a model in a given study does.
\fi

\subsection{Model structure}

For the purpose of illustration, this \thisdoc{} focuses on a three-level tree. Several distinctions below, such as between parameters measured against the root and against the parent, between normalising one nest and normalising a whole level, and the case of sub-nesting within a level, collapse or become invisible in a two-level tree. This has potentially contributed to the confusion in the literature in relation to differences between structures.

The tree used here is made up of a root, upper nests $m_u=1,\dots,M_u$, lower nests $m_l=1,\dots,M_l$, each contained in exactly one upper nest, and elemental alternatives $j$, each contained in exactly one lower nest. Membership is written $m_l\in m_u$ and $j\in m_l$. This balanced form loses no generality. An alternative that sits directly below the root or below an upper nest is placed in degenerate nests with a single member, and an upper nest that is not divided further is given a single lower nest spanning all its alternatives, with a scale equal to its own. In the utility-consistent model, a degenerate nest's parameter cancels from the choice probabilities and is not identified, so its scale too can be set equal to its parent's\footnote{For a degenerate nest $g$ containing only $j$, with structural parameter $\lambda_g$ (Section~\ref{sec:params}), $\lambda_g\log\exp(V_j/\lambda_g)=V_j$, so $\lambda_g$ drops out. In NNNL (Section~\ref{sec:nnnl}) there is no division, so $j$ enters the level above as $\exp(\lambda_g V_j)$ and $\lambda_g$ does not cancel. It multiplies that alternative's utility, and is identified whenever the alternative shares a generic coefficient with others, in which case it measures a difference in sensitivity rather than similarity \citetext{\citealp{hunt2000}; \citealp[Section~5.2]{heiss2002}}. The dummy-node construction of Section~\ref{sec:nnnl} relies on exactly this: a degenerate nest whose parameter is constrained equal to that of a real nest multiplies the utility of the alternative in it by that parameter. \citet[Fig.~1]{koppelman1998a} use this for an alternative that is not nested: placing it in a degenerate nest with the real nest's parameter gives its path the same product as the paths through the real nest. \citet[Section~5.3]{heiss2002} uses it for two real nests with different parameters: each alternative in one nest is placed in a degenerate nest with the other nest's parameter, so that every path carries both parameters, and the products are again equal. In both cases the dummy nest sits directly above a single alternative.}. 

\subsection{Error terms}\label{sec:errors}

Following \citet{williams1977}, the random part of the utility of alternative $j$, in lower nest $m_l$ within upper nest $m_u$, is written as a sum of three independent components, ordered from the root down:
\begin{equation}
\varepsilon_j=\varepsilon_{m_u}+\varepsilon_{m_l|m_u}+\varepsilon_{j|m_l}.
\end{equation}
Here $\varepsilon_{m_u}$ is common to all alternatives in $m_u$, $\varepsilon_{m_l|m_u}$ is common to all alternatives in $m_l$, and $\varepsilon_{j|m_l}$ is specific to $j$. The common components are what make alternatives in the same nest correlated, with variances written as $\omega^2_{m_u}$ and $\omega^2_{m_l|m_u}$.

The root and each nest have a scale, which is the scale of the part of $\varepsilon_j$ that varies \emph{within} them. Within lower nest $m_l$ only $\varepsilon_{j|m_l}$ varies, and is extreme value with scale $\mu_{m_l}$. Within upper nest $m_u$ the component $\varepsilon_{m_u}$ is constant, and what varies is $\varepsilon_{m_l|m_u}+\varepsilon_{j|m_l}$; the construction makes this extreme value with scale $\mu_{m_u}$. At the root nothing is constant, and the whole of $\varepsilon_j$ is extreme value with scale $\mu_0$. Only $\varepsilon_{j|m_l}$ is extreme value in its own right. The common components $\varepsilon_{m_l|m_u}$ and $\varepsilon_{m_u}$ are not, but their distributions are chosen so that adding each to what lies below it again gives an extreme value variable, with a smaller scale \citep{cardell1997}. An extreme value variable with scale $\mu$ has variance $\pi^2/(6\mu^2)$, and the variances of independent components add. At the root and at each nest, what varies splits into the part shared within the nest below and the part that still varies there. As stated above, at the root, what varies is the whole of $\varepsilon_j$, while, within $m_u$, it is $\varepsilon_{m_l|m_u}+\varepsilon_{j|m_l}$. This gives:
\begin{align}
\underbrace{\frac{\pi^2}{6\mu_0^2}}_{\text{varies at the root}}&=\underbrace{\omega^2_{m_u}}_{\text{shared within }m_u}+\underbrace{\frac{\pi^2}{6\mu_{m_u}^2}}_{\text{varies within }m_u}\\
\underbrace{\frac{\pi^2}{6\mu_{m_u}^2}}_{\text{varies within }m_u}&=\underbrace{\omega^2_{m_l|m_u}}_{\text{shared within }m_l}+\underbrace{\frac{\pi^2}{6\mu_{m_l}^2}}_{\text{varies within }m_l}
\end{align}
Variances cannot be negative, so each step down the tree can only remove variance, and the scales cannot decrease as one descends, i.e.:
\begin{equation}
\mu_0\ \leq\ \mu_{m_u}\ \leq\ \mu_{m_l},
\end{equation}
with equality at a step exactly when the shared component there has zero variance, that is, when the nest induces no correlation and could be collapsed into its parent. The unobserved variation remaining within a nest is therefore smallest in the lower nests: alternatives in the same lower nest share more of their unobserved utility than alternatives in different nests, so less unobserved variation separates them.

\subsection{Scales, structural parameters, and anchoring}\label{sec:params}

An important distinction arises between scales and structural parameters, and their use in different descriptions of the model\footnote{In this \thisdoc{}, coefficients on attributes and constants are written with Greek letters other than $\lambda$, $\mu$ and $\sigma$, such as $\beta$, $\theta$, $\alpha$, $\delta$ and $\gamma$, and none of these is used for a scale or a structural parameter. The warning is needed because $\theta$ is the logsum parameter in \citet{daly1987}, in ALOGIT and in several applied guidance documents.}.

\textbf{Scales}, written $\mu$, multiply utility inside an exponential. They increase down the tree and are not bounded by one.

\textbf{Structural parameters} (also called logsum or inclusive value parameters) are ratios of scales, bounded above by one when the conditions below hold. For models with more than two levels, the structural parameters can be anchored in two ways: against the root, written here as $\lambda$, and as implemented in Apollo \citep{hess2019}, or against the parent nest, written here as $\sigma$, and as discussed in \citet[Section~4.3]{train2009}:
\begin{equation}\label{eq:params}
\underbrace{\lambda_{m_u}=\frac{\mu_0}{\mu_{m_u}},\qquad \lambda_{m_l}=\frac{\mu_0}{\mu_{m_l}}}_{\text{root-anchored}}
\qquad\qquad
\underbrace{\sigma_{m_u}=\frac{\mu_0}{\mu_{m_u}},\qquad \sigma_{m_l|m_u}=\frac{\mu_{m_u}}{\mu_{m_l}}=\frac{\lambda_{m_l}}{\lambda_{m_u}}}_{\text{adjacent}}
\end{equation}
At the upper level the parent \emph{is} the root, so $\sigma_{m_u}=\lambda_{m_u}$; the two families separate only from the second level down (i.e. in trees with more than two levels). The root-anchored parameter at any depth is the product of the adjacent ratios along the path, $\lambda_{m_l}=\sigma_{m_u}\,\sigma_{m_l|m_u}$. The consistency conditions are therefore equivalent:
\begin{equation}\label{eq:consistency}
0<\sigma_{m_u}\leq 1 \ \text{ and } \ 0<\sigma_{m_l|m_u}\leq 1 \ \text{ independently}
\qquad\Longleftrightarrow\qquad
0<\lambda_{m_l}\leq\lambda_{m_u}\leq 1 .
\end{equation}
In the adjacent form, each parameter need only lie in $(0,1]$, and no ordering between levels is required. In the root-anchored form, the same conditions appear as an ordering, $\lambda_{m_l}\leq\lambda_{m_u}$. Unconstrained estimation of either form can highlight results that violate these conditions: a $\sigma$ outside $(0,1]$, or a $\lambda$ that exceeds its parent's, the root's being one.\footnote{These conditions guarantee consistency with utility maximisation for all values of the explanatory variables, and are necessary for the error structure of Section~\ref{sec:errors}, which follows the derivation of \citet{williams1977}. They are not necessary for consistency over a limited range of the data. \citet[p.~248]{mcfadden1981} notes that a value above one may be consistent with the GEV conditions over the range of the data, \citet{borschsupan1990} shows this for two-level nested logit, and \citet{herriges1996} derive the resulting bounds, which depend on the probabilities of choosing the nests. \citet{batley2016} show that regularity, which is necessary and sufficient for consistency with utility maximisation in a choice among three alternatives, does not impose the upper bound of one, and report an empirical estimate above one that satisfies it. These local results are established for two-level trees, and \citet[Section~3.5]{carrasco2002} note that admitting values above one may conflict with the ordering between levels in deeper trees. Throughout this \thisdoc{}, consistency with utility maximisation refers to the global conditions.} At two levels, the two forms coincide and the distinction cannot be seen, which is why it is so often missed. Misreading one form as the other misleads in both directions. Read as root-anchored, adjacent values such as $\sigma_{m_u}=0.5$ and $\sigma_{m_l|m_u}=0.8$ appear to violate the ordering, although both satisfy their conditions. Read as adjacent, root-anchored values that do violate it, such as $\lambda_{m_u}=0.5$ and $\lambda_{m_l}=0.8$, appear acceptable, because each lies below one. Apollo \citep{hess2019} stores the root-anchored form, hence its warning that a nest's parameter should not exceed its parent's.

The entire discussion here relates to how the ordering constrains a single root-to-leaf chain. No relation is expected between structural parameters in different branches.

\section{Model structure, implementation and normalisations}\label{sec:model}

This section sets out the nested logit model itself, and the two ways in which it is written in the literature and in software. Section~\ref{sec:ucform} gives the utility-consistent form, in which utilities are divided by the nest parameter within each nest, and shows that the two standard normalisations, at the root and at a lower nest, are one model expressed in different units. Section~\ref{sec:nnnl} gives the undivided form, known as NNNL, which omits that division, and sets out exactly when it coincides with the utility-consistent form. Section~\ref{sec:labels} explains why the labels RU1 and RU2 are an unreliable guide to which form a model uses, and Section~\ref{sec:software} compares how the main packages normalise and report their parameters. The purpose is to separate two questions that are mixed up with each other: which scale is fixed, a choice with no behavioural content, and whether the division is performed, which decides whether the model is consistent with utility maximisation under the specified utilities by construction, or only under the conditions of Section~\ref{sec:nnnl} (Proposition~\ref{prop:nnnl}).

\subsection{The utility-consistent form}\label{sec:ucform}

Following \citet[Section~4.2.3]{train2009}, the deterministic utility of alternative $j$ is split into parts that vary at each level: $W_{m_u}$, common to every alternative in upper nest $m_u$; $W_{m_l}$, common to every alternative in lower nest $m_l$; and $V_{j|m_l}$, specific to $j$. The split is fully general, since any $W$ can be chosen and the remainder assigned to $V_{j|m_l}$. In the divided form of this section the split is a matter of presentation only, since, as shown below, adding a term to $W_{m_l}$ has the same effect on the choice probabilities as adding it to every $V_{j|m_l}$ in the nest. In forms that omit the division (Section~\ref{sec:nnnl}), by contrast, a term common to all alternatives in a nest, such as a nest-specific constant or an attribute of the nest, gives different probabilities depending on whether it is placed in the elemental utilities or specified at the nest level, as some software allows directly \citep[e.g.][]{alogit2022}. Written recursively, and before any normalisation, these composite utilities are:
\begin{equation}\label{eq:composite}
V_{m_l}=W_{m_l}+\frac{I_{m_l}}{\mu_{m_l}},
\qquad
V_{m_u}=W_{m_u}+\frac{I_{m_u}}{\mu_{m_u}},
\end{equation}
where
\begin{equation}\label{eq:logsum}
I_{m_l}=\log\sum_{j\in m_l}\exp\left(\mu_{m_l} V_{j|m_l}\right),
\qquad
I_{m_u}=\log\sum_{m_l\in m_u}\exp\left(\mu_{m_u} V_{m_l}\right)
\end{equation}
are the logsums, or inclusive values, and where, as noted above, $V_{j|m_l}$ is the deterministic utility specific to alternative $j$. Each logsum is the log of the denominator of the choice model within its nest, and $I_{m_l}/\mu_{m_l}$ is the expected maximum utility over the alternatives in $m_l$, up to an additive constant \citep{williams1977}. The logsum terms pass information up the tree.

The choice probabilities at each level are then
\begin{equation}\label{eq:probs}
\begin{gathered}
P(j|m_l)=\frac{\exp\left(\mu_{m_l} V_{j|m_l}\right)}{\sum_{j'\in m_l}\exp\left(\mu_{m_l} V_{j'|m_l}\right)},
\qquad
P(m_l|m_u)=\frac{\exp\left(\mu_{m_u} V_{m_l}\right)}{\sum_{m_l'\in m_u}\exp\left(\mu_{m_u} V_{m_l'}\right)},
\\[2ex]
P(m_u)=\frac{\exp\left(\mu_0 V_{m_u}\right)}{\sum_{m_u'}\exp\left(\mu_0 V_{m_u'}\right)},
\end{gathered}
\end{equation}
such that:
\begin{equation}
	P_j=P(j|m_l)\cdot P(m_l|m_u)\cdot P(m_u)
\end{equation}
This factorisation is a property of the choice probabilities, not a description of behaviour: the model describes a single choice among the elemental alternatives, and the terms in \eqref{eq:probs} are components of that probability rather than separate decisions. For brevity, they are nevertheless referred to as the choice within a lower nest, the choice among lower nests and the choice among upper nests.

The logsum $I_{m_l}$ in \eqref{eq:logsum} is built from utilities multiplied by the nest's own scale $\mu_{m_l}$, and dividing by $\mu_{m_l}$ in \eqref{eq:composite} returns the composite utility $V_{m_l}$ to utility units. The choice among lower nests in \eqref{eq:probs} then multiplies $V_{m_l}$ by the scale of the level above, $\mu_{m_u}$. Given \eqref{eq:composite}, the logsum therefore enters that choice with multiplier $\mu_{m_u}/\mu_{m_l}$, which is the adjacent structural parameter $\sigma_{m_l|m_u}$ of Section~\ref{sec:params}. In the same way, $I_{m_u}$ enters the choice among upper nests with multiplier $\mu_0/\mu_{m_u}=\sigma_{m_u}$. A nest-level term is not divided by its own nest's scale: $W_{m_l}$ enters the choice among lower nests at the scale $\mu_{m_u}$ of the level above, and $W_{m_u}$ enters at scale $\mu_0$. Its coefficients are nonetheless in the same units as those in the elemental utilities: adding a term $w$ to every $V_{j|m_l}$ in nest $m_l$ raises $I_{m_l}$ by $\mu_{m_l}w$, and hence $V_{m_l}$ by $w$, exactly as adding $w$ to $W_{m_l}$ does.

\subsubsection{Identification requirements and normalisation}\label{sec:normalisation}

The probabilities in \eqref{eq:probs} depend on scales and coefficients only through their products. Multiplying every scale by a common factor $c$ and dividing every coefficient by $c$ leaves every probability unchanged, so that only their ratios, and the products of scales with coefficients, are identified.

The tree has $1+M_u+M_l$ scale factors, of which $M_u+M_l$ are identified, as only ratios along a chain from root to leaf are informative\footnote{This assumes every nest has at least two members. A nest with a single member, whether an alternative (a degenerate nest) or a single nest (as when an upper nest that is not divided further is given one lower nest), has a scale that cancels from the choice probabilities and is not identified. One further scale factor is lost for each such nest, and fixing one of these does not count as a normalisation.}. Exactly one scale factor must be fixed, \emph{irrespective of how many levels the tree has}. This is the entire content of ``normalisation'' in what follows, and the point from which much confusion seems to have arisen.

Whichever scale is fixed, the estimated coefficients absorb it. For a single parameter $\theta$, with $\hat\theta$ the value recovered by estimation, $\hat\theta=\mu_0\theta$ if the root scale is fixed (normalisation at the top) and $\hat\theta=\mu_{m_r}\theta$ if the scale of a reference lower nest $m_r$ is fixed (normalisation at the bottom). The notation $\hat V_{j|m_l}$, $\hat W_{m_l}$ and $\hat W_{m_u}$ denotes utilities evaluated at the estimated coefficients, so that $\hat V_{j|m_l}=\mu_0V_{j|m_l}$ in the first case and $\hat V_{j|m_l}=\mu_{m_r}V_{j|m_l}$ in the second, and likewise for $\hat W$. The composite utilities $V_{m_l}$ and $V_{m_u}$ built from them are in the same units, and are not marked separately. The logsums themselves are unaffected by the choice, since $\mu_{m_l}V_{j|m_l}$ is the same whichever scale is fixed.

\paragraph{Normalisation at the top\\} 

Under normalisation at the top, $\mu_0=1$. Since, by \eqref{eq:params}, $\mu_{m_l}=\mu_0/\lambda_{m_l}$ and $\mu_{m_u}=\mu_0/\lambda_{m_u}$, inside each exponential, the utilities are now divided by a root-anchored structural parameter, while the logsums are then multiplied by the same root-anchored structural parameter:
\begin{equation}\label{eq:comp_up}
\begin{gathered}
I_{m_l}=\log\sum_{j\in m_l}\exp\left(\frac{\hat V_{j|m_l}}{\lambda_{m_l}}\right),
\qquad
I_{m_u}=\log\sum_{m_l\in m_u}\exp\left(\frac{V_{m_l}}{\lambda_{m_u}}\right),
\\[1ex]
V_{m_l}=\hat W_{m_l}+\lambda_{m_l}I_{m_l},
\qquad
V_{m_u}=\hat W_{m_u}+\lambda_{m_u}I_{m_u}.
\end{gathered}
\end{equation}
The choice probabilities are then
\begin{equation}\label{eq:probs_up}
\begin{gathered}
P(j|m_l)=\frac{\exp\left(\hat V_{j|m_l}/\lambda_{m_l}\right)}{\sum_{j'\in m_l}\exp\left(\hat V_{j'|m_l}/\lambda_{m_l}\right)},
\\[2ex]
P(m_l|m_u)=\frac{\exp\left(\hat W_{m_l}/\lambda_{m_u}+\left(\lambda_{m_l}/\lambda_{m_u}\right)I_{m_l}\right)}{\sum_{m_l'\in m_u}\exp\left(\hat W_{m_l'}/\lambda_{m_u}+\left(\lambda_{m_l'}/\lambda_{m_u}\right)I_{m_l'}\right)},
\\[2ex]
P(m_u)=\frac{\exp\left(\hat W_{m_u}+\lambda_{m_u}I_{m_u}\right)}{\sum_{m_u'}\exp\left(\hat W_{m_u'}+\lambda_{m_u'}I_{m_u'}\right)}=\frac{\exp\left(V_{m_u}\right)}{\sum_{m_u'}\exp\left(V_{m_u'}\right)}.
\end{gathered}
\end{equation}
With normalisation at the top, the choice among upper-level nests is plain logit in the composite utilities, as the last form of $P(m_u)$ in \eqref{eq:probs_up} shows, with no scale left to multiply them, so all coefficients are in the units of the root. In the choice among lower nests, the lower-nest logsum enters with multiplier $\lambda_{m_l}/\lambda_{m_u}$, which is the adjacent parameter $\sigma_{m_l|m_u}$ of Section~\ref{sec:params} expressed in root-anchored terms, and the nest-level term $\hat W_{m_l}$ is divided by $\lambda_{m_u}$. In the choice within a lower nest, the elemental utilities $\hat V_{j|m_l}$ are divided by $\lambda_{m_l}$, the root-anchored structural parameter of that nest, which is to say multiplied by its scale in root units, $\mu_{m_l}/\mu_0=1/\lambda_{m_l}$; the division is by the root-anchored parameter, not by the adjacent one.

Normalising at the top is the default in Apollo \citep{hess2019,apollomanual} and in Biogeme \citep{biogeme335}, although the two report the same model differently: Apollo gives the root-anchored $\lambda$ used here, Biogeme the scales $\mu=1/\lambda$. Section~\ref{sec:software} compares how the main packages normalise and report their parameters.

\paragraph{Normalisation at the bottom\\} 

Under normalisation at the bottom, $\mu_{m_r}=1$ for one reference lower nest $m_r$ chosen by the analyst. The root scale $\mu_0$ is then free, and the structural parameters keep their definition $\lambda_n=\mu_0/\mu_n$ from Section~\ref{sec:params}. Fixing $\mu_{m_r}$ therefore does not fix $\lambda_{m_r}$: it becomes $\lambda_{m_r}=\mu_0/\mu_{m_r}=\mu_0$, the root scale measured in the units of the reference nest, and it is estimated like every other structural parameter. The parameter released at the root is in exchange for the scale fixed at the bottom, so the parameter count is unchanged. 

Every scale can now be written in terms of structural parameters. Applying the definition to the reference nest gives $\mu_0=\lambda_{m_r}\mu_{m_r}=\lambda_{m_r}$, and applying it to any other nest $n$ then gives $\mu_n=\mu_0/\lambda_n$, so
\begin{equation}
\mu_0=\lambda_{m_r},
\qquad
\mu_{m_u}=\frac{\lambda_{m_r}}{\lambda_{m_u}},
\qquad
\mu_{m_l}=\frac{\lambda_{m_r}}{\lambda_{m_l}}.
\end{equation}
Substituting into \eqref{eq:composite} and the logsums gives
\begin{equation}\label{eq:comp_low}
\begin{gathered}
I_{m_l}=\log\sum_{j\in m_l}\exp\left(\frac{\lambda_{m_r}}{\lambda_{m_l}}\hat V_{j|m_l}\right),
\qquad
I_{m_u}=\log\sum_{m_l\in m_u}\exp\left(\frac{\lambda_{m_r}}{\lambda_{m_u}}V_{m_l}\right),
\\[1ex]
V_{m_l}=\hat W_{m_l}+\frac{\lambda_{m_l}}{\lambda_{m_r}}I_{m_l},
\qquad
V_{m_u}=\hat W_{m_u}+\frac{\lambda_{m_u}}{\lambda_{m_r}}I_{m_u}.
\end{gathered}
\end{equation}
The choice probabilities are then
\begin{equation}\label{eq:probs_low}
\begin{gathered}
P(j|m_l)=\frac{\exp\left((\lambda_{m_r}/\lambda_{m_l})\hat V_{j|m_l}\right)}{\sum_{j'\in m_l}\exp\left((\lambda_{m_r}/\lambda_{m_l})\hat V_{j'|m_l}\right)},
\\[2ex]
P(m_l|m_u)=\frac{\exp\left((\lambda_{m_r}/\lambda_{m_u})\hat W_{m_l}+\left(\lambda_{m_l}/\lambda_{m_u}\right)I_{m_l}\right)}{\sum_{m_l'\in m_u}\exp\left((\lambda_{m_r}/\lambda_{m_u})\hat W_{m_l'}+\left(\lambda_{m_l'}/\lambda_{m_u}\right)I_{m_l'}\right)},
\\[2ex]
P(m_u)=\frac{\exp\left(\lambda_{m_r}\hat W_{m_u}+\lambda_{m_u}I_{m_u}\right)}{\sum_{m_u'}\exp\left(\lambda_{m_r}\hat W_{m_u'}+\lambda_{m_u'}I_{m_u'}\right)}=\frac{\exp\left(\lambda_{m_r}V_{m_u}\right)}{\sum_{m_u'}\exp\left(\lambda_{m_r}V_{m_u'}\right)}.
\end{gathered}
\end{equation}
Comparing the two normalisations shows what the choice of normalisation does and does not change. In the composite utilities, the logsums have different multipliers: $\lambda_{m_l}$ and $\lambda_{m_u}$ under normalisation at the top in \eqref{eq:comp_up}, but $\lambda_{m_l}/\lambda_{m_r}$ and $\lambda_{m_u}/\lambda_{m_r}$ under normalisation at the bottom in \eqref{eq:comp_low}, because the composites are measured in different units. What matters for the probabilities is the multiplier with which each logsum enters the choice at the next level up, and \eqref{eq:probs_up} and \eqref{eq:probs_low} show that this is the same under both: $\lambda_{m_l}/\lambda_{m_u}$ for each lower-nest logsum $I_{m_l}$, one for each lower nest, and $\lambda_{m_u}$ for each upper-nest logsum $I_{m_u}$. These effective multipliers are ratios of scales, $\mu_{m_u}/\mu_{m_l}$ and $\mu_0/\mu_{m_u}$, and so do not depend on which scale is fixed. What changes is the scaling of the utility terms. With normalisation at the top they are in the units of the root: elemental utilities enter the choice within their nest divided by $\lambda_{m_l}$, lower-nest terms enter the choice among lower nests divided by $\lambda_{m_u}$, and upper-nest terms enter the top choice unscaled. With normalisation at the bottom they are in the units of the reference nest, and each of these factors is multiplied by $\lambda_{m_r}$, giving $\lambda_{m_r}/\lambda_{m_l}$, $\lambda_{m_r}/\lambda_{m_u}$ and $\lambda_{m_r}$ respectively. The first of these factors is one in the reference nest itself, since $\mu_{m_r}=1$ is the normalisation, so the elemental utilities there enter unscaled; in every other lower nest it differs from one\footnote{This factor is what keeps normalisation at the bottom consistent with utility maximisation. The undivided form of Section~\ref{sec:nnnl} enters elemental utilities unscaled in every nest, not only in the reference nest as here, and so drops precisely this factor.}. The last forms of $P(m_u)$ in \eqref{eq:probs_up} and \eqref{eq:probs_low} put the difference in one place: the composites enter the top choice unscaled under normalisation at the top, and multiplied by $\lambda_{m_r}$ under normalisation at the bottom.

Both normalisations are consistent with utility maximisation, and they describe the same model: they give the same choice probabilities and the same log-likelihood, and the estimates are related by $\hat\lambda^{\,\text{top}}=\hat\lambda^{\,\text{bottom}}$ for the structural parameters and $\hat\theta^{\,\text{top}}=\hat\lambda_{m_r}\hat\theta^{\,\text{bottom}}$ for the coefficients, since $\hat\theta^{\,\text{top}}=\mu_0\theta$, $\hat\theta^{\,\text{bottom}}=\mu_{m_r}\theta$ and $\mu_0/\mu_{m_r}=\lambda_{m_r}$. Anything that depends only on the probabilities or on ratios of coefficients is therefore the same under both, namely values of time and other willingness-to-pay measures, elasticities, and forecasts. Standard errors and t-ratios of the coefficients themselves are not invariant, however: since $\hat\theta^{\,\text{top}}=\hat\lambda_{m_r}\hat\theta^{\,\text{bottom}}$ involves a second estimate, a coefficient's t-ratio against zero generally differs between the two normalisations. Ratios of coefficients, and their standard errors, are unaffected, since the factor cancels, and so are the estimates and standard errors of the structural parameters, which are the same under both. Otherwise, what differs is only the units of the coefficients, which are those of the root under normalisation at the top, and those of the reference nest under normalisation at the bottom. Normalisation at the bottom also requires an arbitrary choice of reference nest, which changes the units of every coefficient, but nothing else.

\subsection{The undivided form (NNNL)}\label{sec:nnnl}

The undivided form omits the division of utilities by the nest parameter within each nest. It is the form estimated by ALOGIT \citep{daly1987,alogit2022}, whose long use in large-scale transport modelling has made it widely used in practice, and it was the form implemented by the \texttt{nlogit} command in Stata~7.0 \citep{heiss2002}.

\citet[Section~2.1.2]{koppelman1998a} call the form in this section the non-normalised nested logit (NNNL), and the form of the previous section the utility-maximising nested logit (UMNL); \citet{daly2001} also describes it as non-normalised. The present \thisdoc{} argues that that name is misleading. The model is normalised, as it would otherwise not be identified. The difference with the version in the previous section is not one of normalisation, but one of functional form: what the undivided form omits is the division of the utilities by the nest parameter in the within-nest probabilities, at every level. It is therefore a different model, not an un-normalised version of the same one. The NNNL acronym is used here only because it is established in the literature.

The form is often attributed to \citet{daly1987}, which uses it (eq.~15)\footnote{That equation includes nest-level terms, which enter unscaled, with the nest parameter multiplying only the logsum, as in \eqref{eq:logsum_nnnl} and \eqref{eq:probs_nnnl}. \citet[eqs~8--11]{koppelman1998a} set out the two-level form without nest-level terms, noting only, in footnotes, that they may be included, and that in this form doing so is not the same as adding them to each alternative in the nest. The RU1 of \citet{hensher2002} and of NLOGIT treats them differently again (Section~\ref{sec:labels}).}, but that paper set out a method for estimating tree logit models rather than introducing the model, and \citet[p.~719]{daly2001} calls the attribution inaccurate. Both the undivided and the divided form appear in early descriptions of the model \citep{dalyzachary1978,mcfadden1979}, and \citet{daly2001} notes that still earlier presentations of the undivided form were in \citet{benakiva1973}, with closely equivalent forms in the SELNEC models of the late 1960s \citep{wilson1969}. Writing $\tilde V$ and $\tilde W$ for utilities evaluated at the NNNL estimates, and $\tilde I$ and $\tilde\lambda$ for its logsums and multipliers, to keep them apart from the utility-consistent model's $\hat V$, $\hat W$, $I$ and $\lambda$ (the last written $\hat\lambda$ where the two models are set side by side), the form at three levels has logsums
\begin{equation}\label{eq:logsum_nnnl}
\tilde I_{m_l}=\log\sum_{j\in m_l}\exp\left(\tilde V_{j|m_l}\right),
\qquad
\tilde I_{m_u}=\log\sum_{m_l\in m_u}\exp\left(\tilde W_{m_l}+\tilde\lambda_{m_l} \tilde I_{m_l}\right),
\end{equation}
and choice probabilities
\begin{equation}\label{eq:probs_nnnl}
\begin{gathered}
P(j|m_l)=\frac{\exp\left(\tilde V_{j|m_l}\right)}{\sum_{j'\in m_l}\exp\left(\tilde V_{j'|m_l}\right)},
\\[2ex]
P(m_l|m_u)=\frac{\exp\left(\tilde W_{m_l}+\tilde\lambda_{m_l}\tilde I_{m_l}\right)}{\sum_{m_l'\in m_u}\exp\left(\tilde W_{m_l'}+\tilde\lambda_{m_l'}\tilde I_{m_l'}\right)},
\\[2ex]
P(m_u)=\frac{\exp\left(\tilde W_{m_u}+\tilde\lambda_{m_u}\tilde I_{m_u}\right)}{\sum_{m_u'}\exp\left(\tilde W_{m_u'}+\tilde\lambda_{m_u'}\tilde I_{m_u'}\right)}.
\end{gathered}
\end{equation}
%
Comparing \eqref{eq:probs_nnnl} with \eqref{eq:probs_up} term by term shows how the two sets of parameters correspond, taking the utility-consistent model normalised at the top. In both forms each logsum enters the choice at the next level up multiplied by a single factor, but that factor differs: $\hat\lambda_{m_l}/\hat\lambda_{m_u}$ and $\hat\lambda_{m_u}$ in \eqref{eq:probs_up}, and the free parameters $\tilde\lambda_{m_l}$ and $\tilde\lambda_{m_u}$ in \eqref{eq:probs_nnnl}. Matching them gives $\tilde\lambda_{m_u}=\hat\lambda_{m_u}$ and $\tilde\lambda_{m_l}=\hat\lambda_{m_l}/\hat\lambda_{m_u}$: the multipliers of the undivided form are the \emph{adjacent} structural parameters of Section~\ref{sec:params}, and being unrestricted they impose nothing. The difference lies in the utility terms, level by level. In the choice within a lower nest, \eqref{eq:probs_up} uses $\hat V_{j|m_l}/\hat\lambda_{m_l}$ where \eqref{eq:probs_nnnl} uses $\tilde V_{j|m_l}$; in the choice among lower nests, \eqref{eq:probs_up} uses $\hat W_{m_l}/\hat\lambda_{m_u}$ where \eqref{eq:probs_nnnl} uses $\tilde W_{m_l}$; and only in the choice among upper nests are the two treated alike, with $\hat W_{m_u}$ and $\tilde W_{m_u}$ both entering as they are. The utility-consistent form divides each utility term by the structural parameter of the nest that contains it, the root's being one, and the undivided form does not. For the two forms to give the same probabilities, the undivided form's utilities must therefore correspond to $\tilde V_{j|m_l}=\hat V_{j|m_l}/\hat\lambda_{m_l}$, $\tilde W_{m_l}=\hat W_{m_l}/\hat\lambda_{m_u}$ and $\tilde W_{m_u}=\hat W_{m_u}$, so that coefficients on the attributes of alternatives are related by
\begin{equation}\label{eq:coef_nnnl}
\hat\theta_{m_l}=\hat\lambda_{m_l}\,\tilde\theta=\tilde\lambda_{m_u}\tilde\lambda_{m_l}\,\tilde\theta .
\end{equation}
Equivalently, the undivided form restricts coefficients that would otherwise be nest-specific to be proportional to the root-anchored structural parameters, a restriction that Section~\ref{sec:example_tree} sets against that of the utility-consistent form. More generally, the undivided form measures each coefficient in the units of the scale of the level at which its term enters: $\tilde V_{j|m_l}=\mu_{m_l}V_{j|m_l}$, $\tilde W_{m_l}=\mu_{m_u}W_{m_l}$ and $\tilde W_{m_u}=\mu_0W_{m_u}$, so that elemental coefficients are in the units of their own lower nest, coefficients on $W_{m_l}$ in those of the upper nest, and coefficients on $W_{m_u}$ in those of the root. This is why the placement of a term matters in the undivided form but not in the divided one.

The undivided form fixes the scale of every lower nest at one, where normalisation at the bottom fixes only the reference nest's: the factors $\lambda_{m_r}/\lambda_{m_l}$ that normalisation at the bottom keeps on the elemental utilities of the other lower nests, in \eqref{eq:probs_low}, are set to one rather than expressed in the units of the reference nest. Four consequences follow. First, by \eqref{eq:coef_nnnl}, a coefficient $\tilde\theta$ held generic across nests in the undivided form corresponds to a coefficient $\hat\theta_{m_l}$ that differs from one lower nest to another, unless $\hat\lambda_{m_l}$ takes the same value in every lower nest; without nest-level terms, this is the only binding requirement. Second, a coefficient shared between levels, such as a cost coefficient applied both to a nest-level cost and to the alternatives' own costs, is mis-scaled unless $\hat\lambda_{m_l}=\hat\lambda_{m_u}$, that is, unless the lower nest induces no correlation. This is the failure of a coefficient shared across nests, now across levels. Third, and as a consequence of the first, translational invariance is lost: adding a constant to all specified utilities changes the choice probabilities, contrary to the condition of \citet{dalyzachary1978} and \citet{mcfadden1981}, because in the equivalent utility-consistent model the shift is $\hat\lambda_{m_l}c$, which differs between nests. Fourth, and again as a consequence of the first, elasticities behave counter-intuitively \citep[Table~1]{koppelman1998a}: at a common specified coefficient, the direct elasticity of an alternative \emph{inside} a nest is \emph{smaller} than for one outside any nest, so that similarity to other alternatives reduces the effect of a change in an alternative on its own choice probability, where the utility-consistent form makes it larger. In the equivalent utility-consistent model the nested alternative is simply less sensitive, by the factor $\hat\lambda_{m_l}$, and its elasticity is exactly what that model implies.

Given the widespread use of NNNL via ALOGIT in large-scale applied work, the literature has also been concerned with establishing under which conditions NNNL coincides exactly with the utility-consistent form. The two coincide in two main cases. First, they are equivalent when the \emph{product of structural parameters along the path from the root to each elemental alternative is identical}, and, where coefficients are also shared between nest-level terms (i.e.\ in nest-level utilities), when the corresponding partial products, from the root down to the parent of each nest containing such a term, are identical too. Parameters at intermediate levels may then differ freely: $\tilde\lambda_{car}=0.5$ alongside $\tilde\lambda_{PT}=0.8$ is admissible provided the adjacent ratios below them compensate, so that every path has the same cumulative value. Second, they are equivalent when no coefficient is shared between alternatives for which the products of structural parameters along the paths from the root differ, so that the required division is performed implicitly by letting the coefficient differ between those alternatives, allowing analysts to use different sets of structural parameters across paths. \citet[Section~4.3]{heiss2002} shows this second equivalence at two levels, recovering the utility-maximising coefficients by multiplying each NNNL coefficient by its nest parameter, and \citet{daly2001} lists the same cases. The ALOGIT documentation adds a further case in which no dummy nests are needed: alternatives that are never available in the same observations, as when data sets are merged, since the scaling between them then never enters a choice probability \citep[p.~81]{alogit2022}. Outside these cases, Proposition~\ref{prop:nnnl} applies.

\begin{proposition}\label{prop:nnnl}
Consider the undivided form \eqref{eq:logsum_nnnl}--\eqref{eq:probs_nnnl}, with multipliers $\tilde\lambda_{m_u},\tilde\lambda_{m_l}\in(0,1]$.
\begin{enumerate}[label=(\roman*)]
\item Its choice probabilities are those of the utility-consistent form \eqref{eq:probs_up}, normalised at the top, with root-anchored structural parameters $\hat\lambda_{m_u}=\tilde\lambda_{m_u}$ and $\hat\lambda_{m_l}=\tilde\lambda_{m_u}\tilde\lambda_{m_l}$, and with implied utilities
\[
\hat V_{j|m_l}=\hat\lambda_{m_l}\tilde V_{j|m_l},\qquad \hat W_{m_l}=\hat\lambda_{m_u}\tilde W_{m_l},\qquad \hat W_{m_u}=\tilde W_{m_u}.
\]
Since these parameters satisfy \eqref{eq:consistency}, the undivided form is consistent with utility maximisation.
\item The implied utilities have the same coefficients as the specified ones, up to a common scale, if and only if every coefficient is shared only among utility terms with the same factor: the product of multipliers along the path to each elemental alternative, or the partial product for a nest-level term, is equal wherever the coefficient is used.
\item Otherwise, a coefficient specified as generic corresponds, in the implied utilities, to coefficients that differ between nests in proportion to these products. Translational invariance and the expected pattern of elasticities then hold for the implied utilities, but not for the specified ones.
\end{enumerate}
\end{proposition}
\noindent Parts~(i) and~(ii) follow from matching \eqref{eq:probs_up} and \eqref{eq:probs_nnnl} term by term, as above, and part~(iii) follows from part~(ii). The undivided form may therefore be consistent with utility maximisation, but not with the utilities the analyst specified.

In a two-level tree, before any dummy nests are added, the root-to-leaf product \emph{is} the single nest parameter, so the equivalence condition collapses to ``equal structural parameters within the level'', the form in which \citet{koppelman1998a} and \citet{carrasco2002} state it. An alternative placed directly below the root has a path product of one, so if it shares a coefficient with nested alternatives, the condition fails unless the nest parameters are themselves one, which would reduce the model to multinomial logit. The remedy is instead a dummy node: the alternative is placed in a degenerate nest whose parameter is constrained equal to that of the real nest \citep[Fig.~1]{koppelman1998a}, which equalises the products while leaving the tree with two levels. When two real nests have different parameters, by contrast, equality requires the dummy nests of \citet[Section~5.3]{heiss2002}, placed between each real nest and its alternatives, and so a third level.

At three levels, equality at each level remains sufficient but is no longer necessary, because inequality at one level can be offset at another. The dummy-node construction of \citet{koppelman1998a} is stated in exactly the general form, namely adding links so that the product of logsum parameters between the root and each elemental alternative is identical, which is this equivalence condition made constructive. In general, it works by adding a level of degenerate nests between the real nests and the alternatives. For an unnested alternative, as above, a single degenerate nest at the level of the real nests suffices. In the two-level example of \citet[Section~5.3]{heiss2002}, each alternative in the nest `public' is given a degenerate dummy nest whose nesting parameter is constrained to equal that of the nest `other', and each alternative in `other' one whose nesting parameter is constrained to equal that of `public'. Every path then includes the same set of parameters, in a different order, and the parameters of all alternatives are scaled by the same factor $1/(\tilde\lambda_{public}\tilde\lambda_{other})$; only the product matters, not the order. If the dummy nests are added only at the bottom, as in this example, they equalise the products only along the full paths to the elemental alternatives. More generally, a shared coefficient is scaled correctly only if the product of structural parameters between the root and the position in the tree at which it is used is the same wherever it is used, and, for positions at the same level, giving each the same set of structural parameters guarantees this; the ALOGIT documentation states the principle in these terms, adding dummy nests ``so that every alternative is affected by every nesting coefficient'' \citep[p.~81]{alogit2022}. Dummy nests added at the bottom secure it for coefficients in elemental utilities only. A coefficient shared between nest-level utilities in different parts of the tree, or between a nest-level utility and elemental utilities, can therefore remain mis-scaled after they are added. One way to avoid this is to place dummy nests so that every path has the same structural parameters level by level, which may require dummy nests above the real nests as well as below them; every partial product is then the same, and coefficients shared within a level, whether in elemental or in nest-level utilities, are scaled correctly wherever they are used. A coefficient shared across levels, such as a cost coefficient applied both to a nest-level cost and to the alternatives' own costs, cannot be rescued in this way: the product down to the level at which the nest-level term enters always lacks the parameters below that level, so the two uses differ by those parameters unless they equal one (the second consequence above). In the undivided form, such a coefficient has to be specified separately at each level and rescaled after estimation.

One final point deserves attention, namely the inadvertent use of a model whose properties differ from those the analyst intended to specify. Within a single estimated model the division is normally applied at every level or at none. In model systems assembled stage by stage, with a lower stage estimated in one package and its composite utility passed up to higher stages implemented elsewhere, each hand-over is a separate point at which the division can be performed or omitted. Such a system can therefore be utility-consistent at some levels and undivided at others, a possibility that two-level expositions cannot express. The risk is that it is then inconsistent with utility maximisation under the specified utilities, without the analyst being aware of it, since the stage estimated as a single model may well be specified correctly, and the omission sits at a join that no estimation output reports.

The same issue arises for size variables \citep{daly1982}, which measure the number of elementary alternatives in an aggregate alternative, such as a zone, rather than the utility they offer. Whether a given coefficient on a size term delivers proportionality, so that zones of equal utility are chosen in proportion to their size \citep{daly1982}, depends both on whether the utilities are divided and on whether the zones sit in the upper or in the lower nests. Imposing proportionality by fixing that coefficient at one, as is common, can then violate it without this being apparent \citep{hessdaly2026}.

\subsection{A discussion on terminology: RU1, RU2 and NNNL}\label{sec:labels}

Two labels that are widely used when describing NL implementations are RU1 and RU2, which seem to originate in \citet{hensher2002}. They are read in two incompatible ways. For some, RU1 is normalisation at the bottom and RU2 normalisation at the top, two normalisations of the same utility-consistent model. For others, RU1 is the undivided form of Section~\ref{sec:nnnl}, which \citet{koppelman1998a} named the non-normalised nested logit (NNNL), and which is a different model. Both readings have a basis, since the label has been used for both, and this is why it cannot identify a model. In the set-up of \citet{hensher2002} the logsum parameter of a nest is the ratio of two scales, that of the level above to that within the nest, which is the adjacent structural parameter of Section~\ref{sec:params}. Only the ratio is identified, so one of the two must be restricted: RU1 fixes the scale within the nest, and RU2 the scale above it. They present this as a choice of normalisation, emphasising that ``there is one model normalised in different ways'', and define normalisation as ``the process of setting one or more scale parameters equal to unity''. On this reading, the labels belong entirely to the utility-consistent form of Section~\ref{sec:ucform}.

The difficulty lies mainly in the scale that RU1 fixes. At two levels, the scale above every nest is that of the root, and RU2 fixes it. At three levels, a literal reading of the definition would also allow fixing the scale above every lower nest, that is, every upper-nest scale, but this imposes restrictions beyond those needed for identification, and so is not a normalisation at all; it constrains the upper-level structural parameters to be equal, a restriction within the utility-consistent form rather than a departure from it. The reasonable reading of RU2, and the only one consistent with its presentation as a normalisation, is therefore that it fixes the scale at the top of the tree, a single restriction whatever its depth: \emph{normalisation at the top}. RU1 is different. There is one within-nest scale for every lower nest, so ``fixing the scale within the nest'' can mean two different things. Fixing it in one nest is \emph{normalisation at the bottom}: a single restriction, and the same model as RU2, as explained in Section \ref{sec:normalisation}. Fixing it in every nest, while letting the nesting parameters differ across nests, is not possible within the utility-consistent form, where differing nesting parameters imply differing within-nest scales. It leaves the elemental utilities undivided in every nest, which is the undivided form of Section~\ref{sec:nnnl}, a different model. Only when the lower nests share one scale, that is, when their root-anchored parameters are equal, is there a single within-nest scale, in which case the two readings coincide, a case discussed further below. 

The textbook treatment \citep[p.~103]{hensher2015} states the choice in words that allow both readings:
\begin{quote}
\emph{``It is typical to normalize either $\mu_{(j|b)}$ or $\lambda_b$ to 1.0 for one or more of the branches or nests. Normalizing $\mu_{(j|b)}=1.0$ results in models that are said to be normalized to random utility 1 (RU1) while normalizing $\lambda_b=1.0$ produces random utility 2 (RU2) models.''}
\end{quote}
In the quotation, $\mu_{(j|b)}$ is the scale within branch $b$ and $\lambda_b$ the scale of the level above it, not the structural parameter $\lambda$ of this \thisdoc{}.
Normalising $\mu_{(j|b)}$ for one branch is normalisation at the bottom; normalising it for every branch is, with differing nesting parameters, the undivided form. The same passage therefore describes RU1 both as a proper normalisation (as in Section \ref{sec:normalisation}) and, in effect, as NNNL (as in Section~\ref{sec:nnnl}), without distinguishing between the two.

Both readings also appear in \citet{hensher2002}. They introduce RU1 and RU2 as normalisations of one model, but the RU1 probabilities they then write down (their eqs~6--11) have no scale at the lowest level in any branch, while each branch keeps its own inclusive value parameter. With nesting parameters that differ across branches, this is, in the absence of branch-level utility terms, the undivided form of Section~\ref{sec:nnnl}, and not a normalisation. They add that RU1 has been described by others, citing \citet{koppelman1998a} and an unpublished note by \citet{bates1999}, as corresponding to NNNL, and note themselves that ``in this NNNL context \ldots\ there is no explicit scaling in (6) and (7) at the lowest level'' \citep[p.~5]{hensher2002}. \citet{koppelman1998a} do not use the RU1 label; they define NNNL, following \citet{daly1987}, as the form without division, so the identification of RU1 with NNNL is made by \citet{hensher2002} rather than by \citet{koppelman1998a}. 

The results of \citet{hensher2002} are also what make the distinction visible: by estimating each normalisation and specification on the same data, with and without equality constraints, they provide exactly the comparison needed to separate the two questions. Summarised in Table~\ref{tab:hg}, the results show the consequence. With the nesting parameters varying across branches and generic attribute coefficients, RU1 does not reproduce RU2 (Models~2 and~4), as it would if it were a normalisation of the same model, but it does reproduce the undivided form (Model~6), which they describe as ``a typical NL model implemented by many practitioners'' \citep[p.~11]{hensher2002}. That the undivided form fits better here than RU2 ($-184.31$ against $-188.43$) is not evidence in its favour: the two are different models of the same dimension, and neither is nested within the other (Section~\ref{sec:example_tree}). The two have identical coefficients except for that on household income, which enters the utility of one branch only, and which differs between them by exactly that branch's nesting parameter ($0.0172\times2.42=0.0416$). That parameter is RU1's inclusive value parameter for the branch, and at $2.42$ it lies well outside the global conditions of Section~\ref{sec:params}, as do several others in their table. They read the discrepancy with RU2 as ``illustrating the fact that the normalisation has not been handled properly'' (p.~11). Their conclusions keep the same framing: ``normalisation from the bottom (RU1) may cause problems, as the parameter will be scaled differently in each nest'', and ``with the proper normalisation of the model, we see that the NNNL model is, indeed, consistent with RUM'' (p.~15). Since a normalisation cannot change the choice probabilities, or scale a parameter differently in different nests, what these passages describe is, in the terms of this \thisdoc{}, the undivided form. The argument of this \thisdoc{} is that the issue throughout is not one of normalisation, but of whether the division is performed.

\begin{table}[htbp]
\centering
\small
\caption{Log-likelihoods of the models estimated by \citet[Table~1]{hensher2002} on a non-degenerate tree, grouped by value; Model~9 is described in their text rather than their table, and its log-likelihood follows from its stated equivalence to Model~8. There is no Model~7 for this tree: \citet{hensher2002} use that number only for their partially degenerate tree}\label{tab:hg}
\setlength{\tabcolsep}{4pt}
\begin{tabular}{@{}clllr@{}}
\toprule
Model & Specification & Nesting parameters & Attribute coefficients & Log-likelihood \\
\midrule
1 & RU1 & equal across branches & generic & $-190.18$ \\
3 & RU2 & equal across branches & generic & $-190.18$ \\
\midrule
2 & RU1 & varying across branches & generic & $-184.31$ \\
6 & undivided (NNNL) & varying across branches & generic & $-184.31$ \\
\midrule
4 & RU2 & varying across branches & generic & $-188.43$ \\
5 & undivided, with dummy nests & varying across branches & generic & $-188.43$ \\
\midrule
8 & RU1 & varying across branches & alternative-specific & $-177.82$ \\
9 & RU2 & varying across branches & alternative-specific & $-177.82$ \\
\bottomrule
\end{tabular}
\end{table}

The same results show when the two readings coincide. RU1 matches RU2 in two cases, and in both it is still the undivided form: it coincides because these are the cases in which the undivided form is itself utility-consistent. The first is when the nesting parameters are equal across branches (Models~1 and~3). There is then a single within-nest scale, so that fixing it in one nest and fixing it in every nest are the same single restriction, and RU1 is at once a valid normalisation and formally identical to the undivided form. The second is when the nesting parameters vary but every attribute coefficient is alternative-specific (Models~8 and~9), which is the second equivalence condition of Section~\ref{sec:nnnl}. The first case is plausibly where the identification of RU1 with NNNL arose: with a single nesting parameter at the lower level, fixing the lower scale at one and omitting the division is correct, and it becomes the undivided form only when applied to nests with different nesting parameters. At three levels, a single scale at the lower level, which in the undivided form means equal products of the adjacent multipliers down to that level, makes the elemental utilities consistent but not terms entering at the middle level (Section~\ref{sec:nnnl}).

The identification is also built into NLOGIT, which implements the labels of \citet{hensher2002}. The manual presents the nested logit in the undivided form, as ``the `standard' (nonnormalized) model'' \citep[p.~N-500]{greene2016}, and offers three variants, RU1, RU2 and RU3, the last being a reparameterisation of RU2 whose inclusive value parameters are the reciprocals of those under RU2 (p.~N-519). Its RU1 is also undivided at the elemental level, in every branch. It differs from the undivided form of Section~\ref{sec:nnnl} only in that the inclusive value parameter multiplies the branch-level utility terms as well as the logsum, so that without such terms the two give identical results (p.~N-518). With branch-level terms whose coefficients are shared across branches, however, RU1 is a third model, distinct from both the utility-consistent form and the undivided form. Its elemental utilities are undivided, as in the undivided form, but its branch-level utilities are multiplied by the branch's own inclusive value parameter, where the undivided form leaves them unscaled and the utility-consistent form scales them all by the same factor, the scale of the level above. A coefficient shared across branches is therefore scaled differently in each branch at both levels. A branch-level term that enters one branch only is merely rescaled by that branch's inclusive value parameter, so it leaves RU1 and the undivided form the same model, as the income coefficient in Models~2 and~6 shows. This is also the RU1 of \citet{hensher2002}, whose branch-level and limb-level probabilities (their eqs~8 and~10) have this form. Consistently with this, the manual states that RU1 and RU2 are not the same model in general, and are equivalent only when the inclusive value parameters are equal (p.~N-519).

Measured in units, RU1 expresses each term in the units of the scale of the nest to which it belongs, elemental and branch-level utilities in those of the branch and limb-level utilities in those of the limb, where the undivided form expresses each term in the units of the level at which it enters (Section~\ref{sec:nnnl}). At two levels, RU1 is therefore normalisation at the bottom applied separately in every branch: a coefficient shared between a branch-level term and the elemental utilities of the same branch is scaled consistently, as it is not in the undivided form, and only sharing across branches is affected. At three levels this consistency extends to the branch but not to the limb. Both readings of RU1 therefore have a basis in its implementation: within each branch it is normalisation at the bottom, and across branches with differing parameters it behaves as the undivided form.

The labels therefore do not identify a model, and this \thisdoc{} argues that they are best avoided altogether. Only RU2, on its reasonable reading, identifies a single model: the utility-consistent form normalised at the top. RU1 has been used for at least three different things: the utility-consistent form normalised at the bottom, which is the same model as RU2; the undivided form, in which the within-nest scale is fixed in every nest; and the form that \citet{hensher2002} and NLOGIT implement, which is undivided at the elemental level but multiplies nest-level utilities by each nest's own parameter. NNNL is no clearer. The name is inaccurate, since the form is normalised; what distinguishes it is the omitted division (Section~\ref{sec:nnnl}). Its sources also differ in their treatment of nest-level utilities: \citet{daly1987} and NLOGIT's standard model include them unscaled, while \citet{koppelman1998a} leave them out of their formulation, noting only in footnotes that they may be included. The undivided form is sometimes said to resemble normalisation at the bottom, because its coefficients are in the units of the elemental scale, but the resemblance is one of units, not of functional form.

A description of a nested logit model should instead state three things: where the model is normalised, at the top or at the bottom; whether the utilities, at every level, are divided by the structural parameter of the nest that contains them; and whether utilities entering at the level of a nest are in addition multiplied by that nest's own parameter. The first question arises only when the utilities are divided, since the undivided form fixes the scale at the bottom in every nest. Of the three labels, only RU2 answers all three questions, and then only by implication and on its reasonable reading; neither RU1 nor NNNL answers any of them unambiguously. Even under RU2, the reported nesting parameter may be the scale or its reciprocal, the structural parameter (Section~\ref{sec:software}).

\subsection{Software conventions}\label{sec:software}

The same tree estimated in two packages can produce different numbers, ordered differently, both correct. Three things vary: whether the parameter reported is a scale or a structural parameter, whether a structural parameter is anchored at the root or at the parent, and whether the underlying specification divides each utility by the structural parameter of the nest that contains it. Table~\ref{tab:software} sets out how six widely used packages handle each of these. \citet{silberhorn2008} give an earlier comparison of specifications and software. The descriptions refer to the versions listed in the table note. Implementations change over time, and a model's own documentation of the version used remains the authoritative source.

\begin{table}[t!]
\centering
\caption{Specification and reporting of the nested logit model in core software packages}\label{tab:software}
\small
\renewcommand{\arraystretch}{1.35}
\begin{tabular}{@{}l>{\raggedright\arraybackslash}p{2.9cm}>{\raggedright\arraybackslash}p{2.5cm}>{\raggedright\arraybackslash}p{1.8cm}>{\raggedright\arraybackslash}p{4.25cm}@{}}
\toprule
& \textbf{Specification} & \textbf{Reported} & \textbf{Anchor} & \textbf{Well-behaved range} \\
\midrule
Apollo &
divided; $\mu_{root}=1$ &
structural, $\lambda$ &
root &
\mbox{$0<\lambda_{\text{lower}}\leq\lambda_{\text{upper}}\leq 1$} \\
Biogeme &
divided; root scale $=1$ by default, or a nest scale fixed instead &
scale, $\mu=1/\lambda$ &
root &
\mbox{$1=\mu_{\text{root}}\leq\mu_{\text{upper}}\leq\mu_{\text{lower}}$} \\
ALOGIT &
undivided, called ``RU1'' in its own documentation; utility-consistent form via dummy nests &
structural, multiplies each logsum; utilities not divided &
parent &
each in $(0,1]$; no ordering between levels \\
NLOGIT &
undivided (standard form); RU1, RU2, RU3 for two or three levels only &
structural in the standard form, RU1 and RU3; scales under RU2 &
parent in the standard form and RU1; root under RU2 and RU3 &
standard form and RU1: each in $(0,1]$;\newline RU2: \mbox{$1=\mu_{\text{root}}\leq\mu_{\text{upper}}\leq\mu_{\text{lower}}$};\newline RU3: \mbox{$0<\lambda_{\text{lower}}\leq\lambda_{\text{upper}}\leq 1$} \\
Stata &
divided by default; undivided with the \texttt{nonnormalized} option; any number of levels &
structural (dissimilarity parameters) &
root &
\mbox{$0<\lambda_{\text{lower}}\leq\lambda_{\text{upper}}\leq 1$} \\
mlogit (R) &
divided by default; undivided with the \texttt{unscaled} option; one layer of nests &
structural &
root and parent \mbox{(one layer)} &
$0<\lambda\leq 1$ \\
\bottomrule
\end{tabular}

\medskip
\parbox{\linewidth}{\footnotesize\emph{Sources: Apollo 0.3.9 \citep{hess2019,apollomanual}; Biogeme 3.3.5 \citep{bierlaire2023,biogeme335}; ALOGIT 4.5 \citep{alogit2022,daly1987,daly2001}; NLOGIT 6 \citetext{\citealp[Section~N28]{greene2016}; \citealp{hensher2002}}; Stata 19 \citep{stata2025}; mlogit 2.0-0 \citep{croissant2020,mlogit2026}.}}
\end{table}

On the specification, Apollo and Biogeme both estimate the divided, utility-consistent form, normalised at the top by default; Biogeme also allows a nest scale to be fixed instead of the root's. Biogeme's built-in nested logit functions cover one layer of nests below the root, and deeper trees require its generic MEV functions with a user-supplied generating function \citep{biogeme335}. ALOGIT estimates the undivided form, which its own documentation calls ``RU1'' and which \citet{koppelman1998a} call non-normalised; the utility-consistent form is obtained by adding dummy nests. NLOGIT's standard form is also the undivided one, reported with ``No normalizations imposed a priori'' \citep[p.~N-508]{greene2016}. Its RU1, RU2 and RU3 are available only for trees of two or three levels, so a four-level tree can be estimated only in the undivided form, and its RU1 differs from the undivided form in its treatment of nest-level utilities (Section~\ref{sec:labels}).

Stata and the R package \texttt{mlogit} make the choice between the two forms an explicit option. Both estimate the divided form by default, Stata's \texttt{nlogit} command since version~10, having previously estimated the undivided form \citep{heiss2002}, and both offer the undivided form as an option \citep{stata2025,croissant2020}. Stata reports root-anchored parameters, as Apollo does, and allows any number of levels; \texttt{mlogit} allows a single layer of nests, where the two anchors coincide.

On what is reported, Apollo gives structural parameters and Biogeme gives scales, so the same model estimated in the two yields reciprocal numbers. Moving between the two conventions is straightforward, giving $\lambda=1/\mu$ with the root normalised, and the Delta method gives the corresponding standard errors \citep{daly2012}. ALOGIT reports the multiplier of each logsum. NLOGIT reports structural parameters, except under RU2, which its reference guide describes as a scale formulation, with RU3 as its reciprocal \citep[pp.~N-518--519]{greene2016}. The convention appears to have differed between versions (comparing the worked example in the guide, p.~N-522, with Models~1 and~3 of \citealp{hensher2002}), so it is worth confirming which quantity a given version reports before interpreting the estimates.

On the anchor, the parameters of Apollo and Biogeme are measured against the root, as are those of NLOGIT under RU2 and RU3, in the three-level form set out by \citet[eqs~12--17]{hensher2002}. ALOGIT's multipliers, and NLOGIT's in its standard form and under RU1, are instead the adjacent parameters of Section~\ref{sec:params}, each the ratio of parent to child scale \citep[p.~719]{daly2001}. This is why \citet{koppelman1998a} state the dummy-nest fix as a condition on the \emph{product} of logsum parameters from root to leaf, and why the ALOGIT documentation states its consistency condition the same way \citep[pp.~80--82]{alogit2022}. The distinction matters only beyond two levels, where the root-anchored parameter is the product of the adjacent ones along the path, and the adjacent parameter is the ratio of consecutive root-anchored ones.

On the well-behaved range, a reported value above one means opposite things in different packages. In Apollo it signals a violated condition; in Biogeme it is the normal case, because the reported quantity is a scale and scales increase down the tree. Each Apollo parameter must lie in $(0,1]$ \emph{and} not exceed its parent's, and Apollo's diagnostic warns when a nest's value exceeds its parent's. Each ALOGIT parameter must lie in $(0,1]$, but they need not decrease down the tree, since they are adjacent rather than root-anchored. Comparing the two sets directly, without converting, will therefore manufacture an apparent violation or conceal a real one. Where a package reports absolute sensitivities rather than ratios, their magnitude says nothing about consistency at all; only their ratios do.

\section{A worked example}\label{sec:example}

This section applies the preceding results to a three-level tree, to show how the utility-consistent form, under each of the two normalisations, and the undivided form differ in practice. The example is deliberately small, but it has the features that matter: more than one nest at each level below the root, structural parameters that differ between nests at one level, and coefficients that are shared across nests. The structural parameters are set up as in Apollo, measured against the root.

\subsection{Tree, utilities and parameters}\label{sec:example_tree}

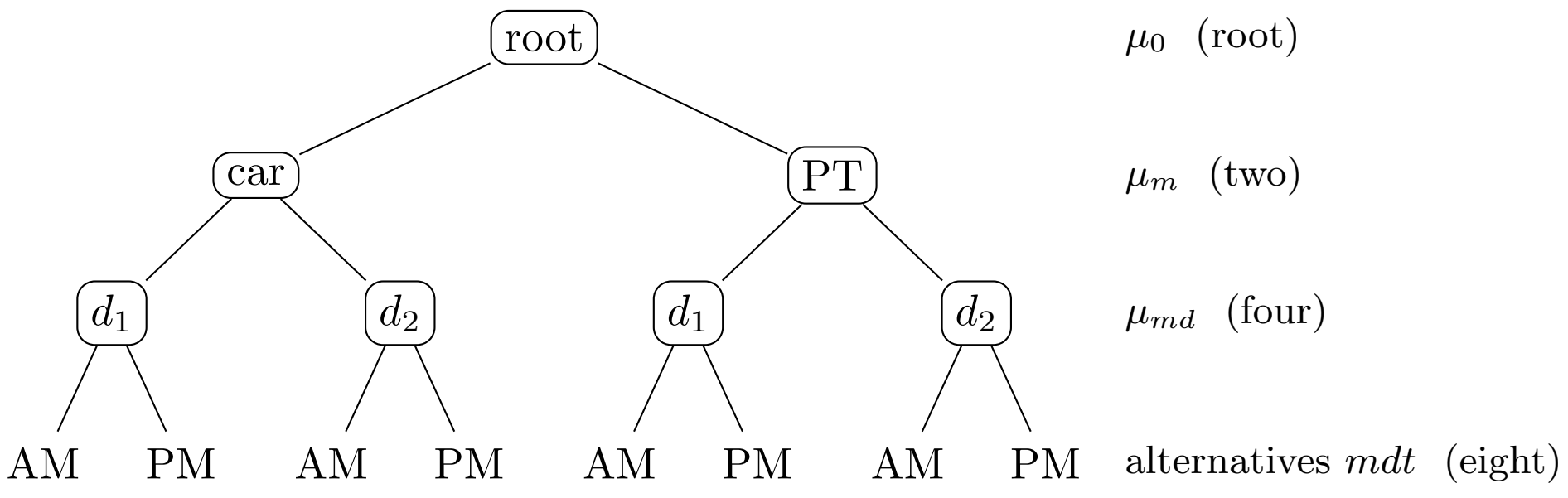
\begin{figure}[htbp]
\centering
\begin{tikzpicture}[
  level 1/.style={sibling distance=46mm, level distance=11mm},
  level 2/.style={sibling distance=23mm, level distance=11mm},
  level 3/.style={sibling distance=11mm, level distance=12mm},
  every node/.style={font=\small},
  nest/.style={draw, rounded corners, inner sep=3pt},
  edge from parent/.style={draw, thin}]
\node[nest] {root}
  child {node[nest] {car}
    child {node[nest] {$d_1$}
      child {node {AM}}
      child {node {PM}}}
    child {node[nest] {$d_2$}
      child {node {AM}}
      child {node {PM}}}}
  child {node[nest] {PT}
    child {node[nest] {$d_1$}
      child {node {AM}}
      child {node {PM}}}
    child {node[nest] {$d_2$}
      child {node {AM}}
      child {node {PM}}}};
\node[anchor=west, font=\footnotesize] at (4.5,0) {$\mu_0$ \ (root)};
\node[anchor=west, font=\footnotesize] at (4.5,-1.1) {$\mu_m$ \ (two)};
\node[anchor=west, font=\footnotesize] at (4.5,-2.2) {$\mu_{md}$ \ (four)};
\node[anchor=west, font=\footnotesize] at (4.5,-3.4) {alternatives $mdt$ \ (eight)};
\end{tikzpicture}
\caption{Tree used in the worked example, with the scale at each level of nests and the number of nests it applies to, and the eight elemental alternatives}\label{fig:example_tree}
\end{figure}

The tree in Figure~\ref{fig:example_tree} has two modes (car and public transport) at the top, two destinations within each mode and two time periods within each destination, giving eight elemental alternatives. Mode is the upper nest ($m_u \equiv m$), destination the lower nest ($m_l \equiv md$), and time period the elemental alternative ($j \equiv t$).

The utility of alternative $mdt$ is
\begin{equation}\label{eq:example_utility}
V_{mdt}=\alpha_m+\delta_t+\beta_{tt,m}\,tt_{mdt}+\beta_{tc}\,tc_{mdt},
\qquad \alpha_{PT}\equiv 0,\quad \delta_{PM}\equiv 0,
\end{equation}
where $\alpha_m$ is a mode-specific constant and $\delta_t$ a time-period constant, the latter common to all modes and destinations. Since only differences in utility matter, one constant of each kind is normalised to zero, here those for public transport and for the PM period. Travel time $tt$ and travel cost $tc$ vary over all three indices, the time coefficient $\beta_{tt,m}$ is mode-specific, and the cost coefficient $\beta_{tc}$ is generic, so that the marginal utility of money is the same in every nest. The mode constant $\alpha_m$ is the same for all alternatives within a mode, so it could also be written as a nest-level utility of the mode nest. In the utility-consistent form the two placements are equivalent, since the composite utilities pass the constant up unchanged. In the undivided form they are not \citep{koppelman1998a}: a constant in the elemental utilities reaches the mode level multiplied by $\tilde\lambda_m\tilde\lambda_d$, while one in a nest-level utility enters unscaled. Since only $\alpha_{car}$ is estimated, the factor is absorbed into it in either placement, which changes its value but not the choice probabilities. The difference would matter for a constant shared across nests, or one whose value is imposed rather than estimated. Here the constant is kept in the elemental utilities, so that there are no nest-level utilities. The generic cost coefficient and time-period constant are what make the example informative, since coefficients shared across nests are the case in which the specifications differ (Section~\ref{sec:nnnl}), while the mode-specific time coefficients provide the contrasting case of a coefficient that is not shared across modes.

The structural parameters are mode-specific at the upper level, $\lambda_{car}$ and $\lambda_{PT}$, and generic at the lower level, with a single $\lambda_d$ shared by all four destination nests. Since the parameters are measured against the root, a generic $\lambda_d$ means that the four destination nests share one scale, $\mu_{md}=\mu_0/\lambda_d$. Global consistency with utility maximisation requires each nest's parameter to lie in $(0,1]$ and not to exceed that of its parent (Section~\ref{sec:params}), so $\lambda_d$ must not exceed the structural parameter of either mode nest, that is, $0<\lambda_d\leq\min(\lambda_{car},\lambda_{PT})\leq 1$. The ratios $\lambda_d/\lambda_{car}$ and $\lambda_d/\lambda_{PT}$ nevertheless differ, so the destination nests are not equally strong relative to their mode: in the mode with the lower $\lambda_m$, more of the substitution caused by a change in one alternative takes place across the mode as a whole, and relatively less is confined to the same destination. Table~\ref{tab:example_params} shows which scale is fixed in each specification and which parameters are then estimated. Each specification estimates eight parameters: five coefficients ($\alpha_{car}$, $\delta_{AM}$, $\beta_{tt,car}$, $\beta_{tt,PT}$ and $\beta_{tc}$) and three structural parameters.

\begin{table}[htbp]
\centering
\caption{Scale normalisation and estimated structural parameters in each specification}\label{tab:example_params}
\small
\begin{tabular}{@{}lp{4.2cm}p{7cm}@{}}
\toprule
& \textbf{Scale fixed} & \textbf{Three structural parameters estimated} \\
\midrule
Top &
$\mu_0=1$ &
$\lambda_{car}$, $\lambda_{PT}$ and $\lambda_d$, measured against the root \\[1ex]
Bottom &
$\mu_{md}=1$, a single restriction, since the four destination nests share one scale &
the same three, with $\lambda_d=\mu_0$ now estimated where the top normalisation fixes $\mu_0$ \\[1ex]
Undivided &
$\mu_{md}=1$ in all four destination nests &
$\tilde\lambda_{car}$, $\tilde\lambda_{PT}$ and a generic $\tilde\lambda_d$, acting as free multipliers of the logsums rather than as ratios of scales \\
\bottomrule
\end{tabular}
\end{table}

The second row answers a natural objection. Normalising at the bottom does not reduce the number of parameters: fixing the destination scale releases $\mu_0$, whose value relative to it, $\lambda_d$, is then estimated. With a single destination scale, fixing it in one nest and fixing it in all four are the same restriction, which is the case in which the two readings of RU1 coincide (Section~\ref{sec:labels}).

The third row is the more important. The undivided form, with a generic multiplier on the destination logsums as an analyst would naturally specify it, has the same number of parameters as the utility-consistent form. It is not a restricted version of that model, which a likelihood ratio test could reject, but a different model of the same dimension. This is why it can be estimated without difficulty and report plausible parameter values, with nothing in the output to show that it is not the utility-consistent model.

The two can nevertheless be compared formally, since both are restrictions of a common, more general model: the utility-consistent form in which the coefficients, and the structural parameters of the destination nests, are free to differ by mode. This model has eleven parameters: $\alpha_{car}$, $\beta_{tt,car}$ and $\beta_{tt,PT}$ as before, mode-specific $\beta_{tc,m}$ and $\delta_{AM,m}$, and four structural parameters $\lambda_{car}$, $\lambda_{PT}$, $\lambda_{car,d}$ and $\lambda_{PT,d}$, all measured against the root. The utility-consistent specification of Table~\ref{tab:example_params} imposes three restrictions on it: $\beta_{tc,car}=\beta_{tc,PT}$, $\delta_{AM,car}=\delta_{AM,PT}$ and $\lambda_{car,d}=\lambda_{PT,d}$. The undivided form imposes three different ones. By \eqref{eq:coef_nnnl}, its generic coefficients tie the ratios of the mode-specific coefficients to the ratio of the structural parameters, $\beta_{tc,car}/\beta_{tc,PT}=\delta_{AM,car}/\delta_{AM,PT}=\lambda_{car,d}/\lambda_{PT,d}$, and its generic multiplier on the destination logsums makes the structural parameters of the destination nests generic relative to the parent rather than the root, $\lambda_{car,d}/\lambda_{car}=\lambda_{PT,d}/\lambda_{PT}$. Each specification can therefore be tested against the general model with a likelihood ratio test on three degrees of freedom, but neither is nested within the other.

The general model serves only as a diagnostic device, not as a proposed specification: an analyst would rarely want the marginal utility of money to differ between modes, which is why the cost coefficient in \eqref{eq:example_utility} is generic. Nor would a result favouring the restrictions of the undivided form make that form the preferred model. Those restrictions make the implied cost sensitivities proportional to the structural parameters, so that the correlation structure dictates how money is valued, which has no behavioural motivation. What such a result shows is that the generic specification is rejected by the data, and the appropriate response is to specify, in the utility-consistent form, whatever differences in sensitivity can be justified on behavioural grounds, and to estimate them as such. The same holds in general: with nest-specific coefficients $\theta_{m_l}$ on the attributes of alternatives, the utility-consistent form with generic coefficients sets $\theta_{m_l}=\theta$, while the undivided form sets $\theta_{m_l}=\lambda_{m_l}\tilde\theta$, holding the implied coefficients proportional to the model's own root-anchored structural parameters rather than equal.

\subsection{Probabilities and logsums}\label{sec:example_levels}

As before, $\hat V_{mdt}$ denotes the utility at the estimates of the utility-consistent model, in the units of the normalisation used, and $\tilde V_{mdt}$ and $\tilde\lambda$ the utility and the multipliers at the estimates of the undivided form, as in Section~\ref{sec:nnnl}. Since there are no nest-level utilities, each nest enters the choice at the level above only through its logsum, $I_{md}$ or $I_m$ in the utility-consistent form and $\tilde I_{md}$ or $\tilde I_m$ in the undivided form, multiplied as in \eqref{eq:probs_up}, \eqref{eq:probs_low} and \eqref{eq:probs_nnnl}. Table~\ref{tab:example_probs} sets out the resulting probabilities.

\begin{table}[htbp]
\centering
\caption{Choice probabilities and logsums in the worked example}\label{tab:example_probs}
\footnotesize
\renewcommand{\arraystretch}{2.9}
\setlength{\tabcolsep}{4pt}
\begin{tabular}{@{}lccc@{}}
\toprule
& \textbf{Top} ($\mu_0=1$) & \textbf{Bottom} ($\mu_{md}=1$) & \textbf{Undivided} \\
\midrule
$P(t\mid m,d)$ &
$\dfrac{e^{\hat V_{mdt}/\lambda_{d}}}{\sum_{t'}e^{\hat V_{mdt'}/\lambda_{d}}}$ &
$\dfrac{e^{\hat V_{mdt}}}{\sum_{t'}e^{\hat V_{mdt'}}}$ &
$\dfrac{e^{\tilde V_{mdt}}}{\sum_{t'}e^{\tilde V_{mdt'}}}$ \\
logsum of $md$ &
$I_{md}=\log\sum_{t'}e^{\hat V_{mdt'}/\lambda_{d}}$ &
$I_{md}=\log\sum_{t'}e^{\hat V_{mdt'}}$ &
$\tilde I_{md}=\log\sum_{t'}e^{\tilde V_{mdt'}}$ \\
$P(d\mid m)$ &
$\dfrac{e^{(\lambda_d/\lambda_{m})I_{md}}}{\sum_{d'}e^{(\lambda_d/\lambda_{m})I_{md'}}}$ &
$\dfrac{e^{(\lambda_d/\lambda_{m})I_{md}}}{\sum_{d'}e^{(\lambda_d/\lambda_{m})I_{md'}}}$ &
$\dfrac{e^{\tilde\lambda_{d}\tilde I_{md}}}{\sum_{d'}e^{\tilde\lambda_{d}\tilde I_{md'}}}$ \\
logsum of $m$ &
$I_{m}=\log\sum_{d'}e^{(\lambda_d/\lambda_{m})I_{md'}}$ &
$I_{m}=\log\sum_{d'}e^{(\lambda_d/\lambda_{m})I_{md'}}$ &
$\tilde I_{m}=\log\sum_{d'}e^{\tilde\lambda_{d}\tilde I_{md'}}$ \\
$P(m)$ &
$\dfrac{e^{\lambda_m I_{m}}}{\sum_{m'}e^{\lambda_{m'} I_{m'}}}$ &
$\dfrac{e^{\lambda_m I_{m}}}{\sum_{m'}e^{\lambda_{m'} I_{m'}}}$ &
$\dfrac{e^{\tilde\lambda_{m}\tilde I_{m}}}{\sum_{m'}e^{\tilde\lambda_{m'}\tilde I_{m'}}}$ \\
\bottomrule
\end{tabular}
\end{table}

The table shows, first, that the division takes place at every level, not only at the bottom. Reading down the first column, each utility is divided by the structural parameter of the nest that contains it, so that the logsum of a nest enters the probability at the level above multiplied by the ratio of its own parameter to that of its parent: $\lambda_d/\lambda_m$ in $P(d\mid m)$, and $\lambda_m$ in $P(m)$, the root's parameter being one. The undivided form omits the division throughout, and its parameters appear only as free multipliers of logsums. A two-level presentation cannot show this, since the root parameter is one and the division at the top is therefore invisible.

At the lowest level, the forms agree. Because the destination scale is common, the elemental probabilities under normalisation at the bottom are undivided, exactly as in the undivided form, and under normalisation at the top they are divided by the same $\lambda_d$ in every nest, which only changes the units of the coefficients. In terms of the equivalence conditions of Section~\ref{sec:nnnl}, the product of the adjacent parameters along the path from the root to each elemental alternative, $\lambda_m\cdot(\lambda_d/\lambda_m)$, equals $\lambda_d$ in both modes, since $\lambda_m$ cancels.

The forms start to differ in the row for $P(d\mid m)$. In the utility-consistent form, under either normalisation, the logsum of each destination nest enters the choice of destination with the multiplier $\lambda_d/\lambda_m$, which is common to the destinations within a mode but differs between the two modes. In the undivided form, it enters with the single multiplier $\tilde\lambda_d$ in both modes. Matching the two would require $\tilde\lambda_d=\lambda_d/\lambda_{car}$ and $\tilde\lambda_d=\lambda_d/\lambda_{PT}$ at once, which is impossible when $\lambda_{car}\neq\lambda_{PT}$. A structural parameter for the destination nests that is generic when measured against the root, as in Apollo, is therefore mode-specific when measured against the parent, which is how the logsum multipliers of the undivided form are measured. Conversely, a generic multiplier on the destination logsums in the undivided form corresponds to mode-specific root-anchored structural parameters, $\lambda_{md}=\tilde\lambda_m\tilde\lambda_d$.

The failure shows up one level higher, in the row for $P(m)$. If a constant $c$ is added to every elemental utility, every term of $P(d\mid m)$ gains the same factor within a mode in all three columns, so the destination probabilities are unchanged. In the utility-consistent form, the exponent $\lambda_m I_m$ of $P(m)$ then increases by the same amount for car as for public transport, so $P(m)$ is unchanged; that amount is $c$ under the top normalisation and $\lambda_d c$ under the bottom one. In the undivided form, $\tilde I_m$ increases by $\tilde\lambda_d c$, so the numerator of $P(m)$ gains $e^{\tilde\lambda_m\tilde\lambda_d c}$, which differs between the two modes unless $\tilde\lambda_{car}=\tilde\lambda_{PT}$. A uniform shift in the specified utilities therefore changes the mode shares, contrary to the condition of \citet{dalyzachary1978}, because in the equivalent utility-consistent model it is not uniform.

The first two columns differ only at the lowest level, and this is a consequence of the structural parameter of the destination nests being generic. Normalising at the bottom then removes the division by $\lambda_d$ in every destination nest, so that the coefficients are measured in units of the common destination scale, larger by the factor $1/\lambda_d$ than under normalisation at the top. With structural parameters that differed between destination nests, normalisation at the bottom would fix the scale of one reference nest only, and the utilities in the other nests would still be multiplied by $\lambda_{m_r}/\lambda_{md}$, where $m_r$ is the reference nest, as in \eqref{eq:probs_low}.

\subsection{Assumptions implied by the undivided form}\label{sec:example_assumes}

The implications of the undivided form are clearest through the coefficient relation \eqref{eq:coef_nnnl}, which here reads $\hat\beta_{m}=\tilde\lambda_{m}\tilde\lambda_d\,\tilde\beta$. A single generic cost coefficient $\tilde\beta_{tc}$ estimated in the undivided form corresponds to a utility-consistent model in which the marginal utility of money is $\tilde\lambda_{car}\tilde\lambda_d\tilde\beta_{tc}$ for car trips and $\tilde\lambda_{PT}\tilde\lambda_d\tilde\beta_{tc}$ for public transport trips. Travellers are thus assumed to value money differently according to the mode they use, without anything in the specification to motivate it. The same applies to the time-period constant, so that the implied preference for the morning peak also differs by mode. The mode-specific time coefficients, by contrast, impose no such restriction, since they are free to absorb the factor for their mode. This relies on the factor varying with mode alone, which here follows from the structure of the tree: the mode nests are the only nests whose multipliers differ, and the multiplier on the destination logsums below them is generic. Had the products also differed between destination nests, a mode-specific coefficient would no longer have absorbed them. Values of time within a mode are unaffected by reading the undivided estimates as a utility-consistent model, because all coefficients in a mode are rescaled by the same factor. They need not, however, equal those of the utility-consistent model with a generic cost coefficient, which imposes a different restriction and would be estimated separately (Section~\ref{sec:example_tree}). What the undivided form distorts is the comparison, between modes, of the coefficients that are shared.

To reproduce the utility-consistent model with a generic $\lambda_d$, the undivided form would need a separate multiplier on the destination logsums in each mode, $\tilde\lambda_{m,d}=\lambda_d/\lambda_m$, constrained so that the products are equal, $\tilde\lambda_{car}\tilde\lambda_{car,d}=\tilde\lambda_{PT}\tilde\lambda_{PT,d}$. This leaves three free multipliers, as in the utility-consistent form. The constraint is needed, as, estimated freely, the separate multipliers would in general give unequal products again, and the generic coefficients would remain mis-scaled. In software that allows multipliers only to be free or fixed, it can be imposed by writing $\tilde\lambda_{PT,d}=\tilde\lambda_{car}\tilde\lambda_{car,d}/\tilde\lambda_{PT}$, or equivalently with dummy nests, as in Section~\ref{sec:nnnl}. Estimating $\beta_{tc}$ and $\delta_{AM}$ separately for each mode would satisfy the second equivalence condition of Section~\ref{sec:nnnl}, but only by abandoning the generic cost coefficient: the separate estimates would combine any genuine difference in cost sensitivity between modes with the factors $\tilde\lambda_m\tilde\lambda_d$, and the two could be separated only by rescaling after estimation. It is the combination of generic coefficients with unequal products that fails. With structural parameters that also differ between destination nests, the products differ at the elemental level too, and the marginal utility of money would vary by destination nest as well as by mode.

Table~\ref{tab:example_tc} makes these points numerically. It uses illustrative attributes for the eight alternatives, listed in the table note, and, in root units, $\beta_{tc}=-0.15$ per pound, $\beta_{tt,car}=-0.035$ and $\beta_{tt,PT}=-0.025$ per minute, so that the values of time are \pounds14 and \pounds10 per hour, with $\alpha_{car}=0.3$ and $\delta_{AM}=0.1$. The car constant is set separately in each specification so that all specifications give the same base car share, and the table reports the car share after \pounds1 is added to the cost of every alternative, and after \pounds1 is added to car trips to the first destination only. In panel~A, the utility-consistent model has root-anchored structural parameters $\lambda_{car}=0.8$ and $\lambda_{PT}=0.5$ for the mode nests and a generic $\lambda_d=0.4$ for the destination nests. By \eqref{eq:params}, the adjacent structural parameters of the destination nests are $\lambda_d/\lambda_m$, that is $0.4/0.8=0.5$ under car and $0.4/0.5=0.8$ under public transport, and these are the multipliers that the undivided form needs on the destination logsums (Section~\ref{sec:nnnl}). In the undivided form, the specified coefficients are the root-unit ones divided by $\lambda_d$, so that the cost coefficient is $-0.15/0.4=-0.375$ in both modes. With a generic multiplier $\tilde\lambda_d=0.6$ on the destination logsums, the implied root-anchored structural parameters of the destination nests are the products $\tilde\lambda_m\tilde\lambda_d$, $0.8\times0.6=0.48$ under car and $0.5\times0.6=0.30$ under public transport, and by \eqref{eq:coef_nnnl} the implied cost coefficients are $0.48\times(-0.375)=-0.18$ and $0.30\times(-0.375)=-0.1125$. The values of time within each mode remain \pounds14 and \pounds10, but the uniform charge, which cannot matter under the specified utilities, moves $1.7$ percentage points out of car. With dummy nests, which impose the multipliers $0.5$ and $0.8$ on the destination logsums, both paths have the product $0.4$, and the undivided form reproduces the utility-consistent model exactly. In panel~B, the mode nests share one structural parameter, $0.65$, and the undivided form has the multiplier $0.6$ on the destination logsums, so that the root-anchored structural parameter of the destination nests is $0.65\times0.6=0.39$ under both modes. The two forms are then the same model: they agree in every column, and neither responds to the uniform charge. The agreement is bought by restricting the correlation structure, which shows in the response to the targeted charge ($53.10$ against $53.24$ per cent), and, estimated on data generated by the model of panel~A, the restricted model would also fit worse.

\begin{table}[t!]
\centering
\caption{Calculated example for the tree of Figure~\ref{fig:example_tree}: implied cost coefficients in root units and car share under each specification}\label{tab:example_tc}
\small
\begin{tabular}{@{}>{\raggedright\arraybackslash}p{6.4cm}rrrrr@{}}
\toprule
& \multicolumn{2}{c}{Implied cost coefficient} & \multicolumn{3}{c}{Car share (\%)} \\
\cmidrule(lr){2-3}\cmidrule(l){4-6}
Specification & car & PT & base & \pounds1 on all & \pounds1 on car to $d_1$ \\
\midrule
\multicolumn{6}{@{}l}{\emph{A. Mode nests with different structural parameters} ($\lambda_{car}=0.8$, $\lambda_{PT}=0.5$)}\\
utility-consistent, generic $\lambda_d=0.4$ & $-0.1500$ & $-0.1500$ & 55.74 & 55.74 & 53.24 \\
undivided, generic $\tilde\lambda_d=0.6$ & $-0.1800$ & $-0.1125$ & 55.74 & 54.06 & 52.60 \\
undivided, with dummy nests & $-0.1500$ & $-0.1500$ & 55.74 & 55.74 & 53.24 \\
\midrule
\multicolumn{6}{@{}l}{\emph{B. Restricted: mode nests with a common structural parameter}}\\
utility-consistent, $\lambda_m=0.65$, $\lambda_d=0.39$ & $-0.1500$ & $-0.1500$ & 55.74 & 55.74 & 53.10 \\
undivided, $\tilde\lambda_m=0.65$, $\tilde\lambda_d=0.6$ & $-0.1500$ & $-0.1500$ & 55.74 & 55.74 & 53.10 \\
\bottomrule
\end{tabular}

\medskip
\parbox{\linewidth}{\footnotesize Attributes (time in minutes, cost in pounds), for $d_1$ AM, $d_1$ PM, $d_2$ AM, $d_2$ PM: car $(30,6)$, $(25,5)$, $(40,8)$, $(35,7)$; public transport $(45,4)$, $(40,3.5)$, $(50,5)$, $(48,4.5)$. Values of time \pounds14 and \pounds10 per hour in every specification. Calculated, not estimated.}
\end{table}

\subsection{The same tree with a generalised cost}\label{sec:example_gc}

Suppose now that the time and cost terms are replaced by a generalised cost expressed in units of time,
\begin{equation}\label{eq:example_gc}
G_{mdt}=tt_{mdt}+\frac{tc_{mdt}}{v_m},\qquad V_{mdt}=\alpha_m+\delta_t-\gamma\,G_{mdt},
\end{equation}
where $v_m$ is an externally supplied value of time, possibly specific to the mode. Such hierarchies are usually written in terms of a sensitivity to generalised cost at each level, $\kappa$, which combines the scale at that level with the marginal utility of a generalised minute\footnote{Guidance and software commonly write $\lambda$ for this sensitivity, which is a scale times a coefficient, and not the structural parameter $\lambda$ of this \thisdoc{}.}. Composite costs are then formed in one of two ways. In the divided form, the logsum over destinations is converted back to generalised minutes, $G^{comp}_{m}=-\kappa_{d,m}^{-1}\ln\sum_d\exp(-\kappa_{d,m}G_{md})$, and enters the choice of mode multiplied by the sensitivity of that choice, $\kappa_M$. In the undivided form, the logsum is kept in the units of the level below and multiplied by a scaling parameter, $\theta\ln\sum_d\exp(-\kappa_{d,m}G_{md})$. The two agree when the destination sensitivity is common to the modes, with $\theta=\kappa_M/\kappa_d$ the adjacent structural parameter. When it differs by mode, they agree only if the scaling parameter differs too, $\theta_m=\kappa_M/\kappa_{d,m}$.

With a single $\theta$ and mode-specific $\kappa_{d,m}$, every path has the same product of structural parameters and the coefficient on generalised cost is mode-specific, so the undivided form meets the equivalence conditions of Section~\ref{sec:nnnl}, but only under utilities in which the $\kappa_{d,m}$ differ through the marginal utility of a generalised minute rather than through scale. The choice of mode then has sensitivity $\theta\kappa_{d,m}$, and since the values of time tie money to time, the marginal utility of money differs between modes by more than they imply, at both levels. Unlike the mode-specific time coefficients of Section~\ref{sec:example_assumes}, no free coefficient can absorb this. If the $\kappa_{d,m}$ instead differ through unobserved variation, as describing $\theta$ as a ratio of sensitivities suggests, only $\theta_m$ reproduces the divided form.

\ifacademic
Both forms appear in the UK Department for Transport's guidance on variable demand modelling, TAG Unit M2.1 \citep{dft2025}. It forms its composite cost in the divided form (its Section~3.2.1), while its incremental formulation (its Appendix~D), its reference implementation \citep{diadem2026}, and the models from which its illustrative parameters are drawn \citep{mva2005} use the undivided form. Its illustrative values show that the difference is not small. The destination choice sensitivities are mode-specific, for home-based work $0.065$ per generalised minute by car and $0.033$ by public transport (its Table~3), while the main mode scaling parameter is a single value, $0.70$ (its Table~4), described as reflecting ``the ratios of the lambdas'' (its Appendix~D.1.1). Applied in the undivided form, they give the choice of mode a sensitivity of about $0.046$ to a generalised minute by car and $0.023$ by public transport, and ratios between $1.9$ and $2.5$ across the purposes tabulated. The source models report the corresponding ``different parameters for car and pt'' without comment \citep[Section~1.6]{mva2005}. Whenever the destination sensitivities differ by mode, the two formulations in the same guidance are therefore different models. The guidance also recommends a split between public transport sub-modes below the main mode choice (its Section~4.7.10), which its reference implementation leaves to a separate model, an instance of the problems discussed next.
\fi

Table~\ref{tab:example_gc} repeats the exercise for the generalised cost formulation, on the choice of mode and destination, using the morning attributes of Table~\ref{tab:example_tc} and values of time of \pounds14 and \pounds10 per hour by car and public transport. Since the values of time differ, a pound is a different number of generalised minutes by each mode, so a uniform charge should change the shares in every specification. The test of translational invariance is instead the addition of the same number of generalised minutes, here five, to every alternative. In panel~A, the sensitivities to generalised cost in the choice of destination are $\kappa_{d,car}=0.06$ and $\kappa_{d,PT}=0.03$, and the divided form has a single sensitivity in the choice of mode, $\kappa_M=0.024$. Each $\kappa$ is a scale times the same coefficient on a generalised minute, so the ratio of the sensitivities at adjacent levels is the ratio of the scales, and hence the adjacent structural parameter of the destination nests: $\theta_m=\kappa_M/\kappa_{d,m}$, that is $0.024/0.06=0.4$ under car and $0.024/0.03=0.8$ under public transport. The undivided form with a single $\theta=0.6$ instead gives the choice of mode a sensitivity $\theta\kappa_{d,m}$ of $0.6\times0.06=0.036$ to a generalised minute by car and $0.6\times0.03=0.018$ by public transport, and loses $2.25$ points of car share when every alternative becomes five minutes longer. It also gets the direction of the response to a uniform charge wrong: because a pound is fewer generalised minutes by car, the car share should rise, as it does in the divided form, from $49.98$ to $51.00$ per cent, but it falls to $48.82$ per cent. With $\theta_m=0.4$ and $0.8$, the equivalent of dummy nests, the undivided form again reproduces the divided one. Panel~B shows the restricted case, a common sensitivity of $0.045$ in the choice of destination, so that the destination nests share one scale. The single $\theta=0.6$ of the undivided form then corresponds to $\kappa_M=0.6\times0.045=0.027$ in the divided form, and the two forms coincide, at the cost of a restricted correlation structure that shows in the response to the charge.

\begin{table}[t!]
\centering
\caption{Calculated example with a generalised cost: sensitivity of the choice of mode to a generalised minute and car share under each specification}\label{tab:example_gc}
\small
\begin{tabular}{@{}>{\raggedright\arraybackslash}p{6.4cm}rrrrr@{}}
\toprule
& \multicolumn{2}{c}{Mode-choice sensitivity} & \multicolumn{3}{c}{Car share (\%)} \\
\cmidrule(lr){2-3}\cmidrule(l){4-6}
Specification & car & PT & base & $+5$ gen.\ min & \pounds1 on all \\
\midrule
\multicolumn{6}{@{}l}{\emph{A. Sensitivities in the choice of destination differing by mode} ($\kappa_{d,car}=0.06$, $\kappa_{d,PT}=0.03$)}\\
divided & $0.024$ & $0.024$ & 49.98 & 49.98 & 51.00 \\
undivided, single $\theta=0.6$ & $0.036$ & $0.018$ & 49.98 & 47.73 & 48.82 \\
undivided, $\theta_m=\kappa_M/\kappa_{d,m}$ (dummy nests) & $0.024$ & $0.024$ & 49.98 & 49.98 & 51.00 \\
\midrule
\multicolumn{6}{@{}l}{\emph{B. Restricted: common sensitivity in the choice of destination}}\\
divided, common $\kappa_d=0.045$, $\kappa_M=0.027$ & $0.027$ & $0.027$ & 49.98 & 49.98 & 51.13 \\
undivided, common $\kappa_d=0.045$, $\theta=0.6$ & $0.027$ & $0.027$ & 49.98 & 49.98 & 51.13 \\
\bottomrule
\end{tabular}

\medskip
\parbox{\linewidth}{\footnotesize Morning attributes of Table~\ref{tab:example_tc}; values of time \pounds14 and \pounds10 per hour; the car constant is set in each specification to give the same base car share. Calculated, not estimated.}
\end{table}

More generally, wherever a hierarchy built on a generalised cost has two or more nests at one level with different sensitivities, whether through sub-nesting within a level, or through one choice repeated under several upper nests with a separate sensitivity in each, the division, or dummy nests that reproduce it, is all that stands between the model and the behaviour described above. A split between public transport sub-modes below the main mode choice, computed in a separate model, combines both sub-nesting and the hand-over between stages of Section~\ref{sec:nnnl}.

\FloatBarrier
\section{Conclusions}\label{sec:conclusions}

This paper has sought to clarify the relationship between the different versions of the nested logit model, their normalisations and the labels used to describe them. Its central point can be put briefly: normalisation changes the units, whereas division changes the model. Normalising at the top or at the bottom gives the same model, with the same probabilities and the same fit, and differs only in the units in which the coefficients and structural parameters are reported. Normalisation at the top is the simplest to implement and to interpret, and should in general be preferred. The exception is sequential estimation, in which the lower levels are estimated first, so that the scale is necessarily fixed at the bottom. If coefficients are shared across the lower nests and estimated jointly at that stage, the scale is in fact fixed in every lower nest at once, and the model then represents the specified utilities only if the root-anchored structural parameters of those nests are equal, so that they share one scale. 

These points concern normalisation only. What changes the model is whether each utility is divided by the structural parameter of the nest that contains it. The undivided form omits this division. It has been described as non-normalised, but misleadingly so: it fixes the scale in every lower nest, and so is normalised, and it has the same number of parameters as the utility-consistent form. It is simply a different model, and it coincides with the utility-consistent form only when the products of structural parameters along the paths from the root are equal wherever a coefficient is shared, or when no coefficient is shared between paths whose products of structural parameters differ. Outside these cases it is consistent with utility maximisation only under utilities other than those specified (Proposition~\ref{prop:nnnl}): a coefficient that is specified as generic becomes, in the equivalent utility-consistent model, specific to the nests with different parameters, as the worked example shows for the marginal utility of money, and translational invariance and the expected pattern of elasticities are lost with respect to the specified utilities. The example also shows that a parameter that is generic when measured against the root is not generic when measured against the parent, so that the same verbal specification can describe different models in different software. A generalised cost formulation, in which the trade-offs between attributes are fixed externally, leaves no coefficient that could absorb the factors implied by the undivided form without giving up the common sensitivity to generalised cost. Unless the products of structural parameters are equal along every path, because the sensitivities agree within each level or because dummy nests make them agree, such a hierarchy represents the specified utilities only if it performs the division. The same distinction affects size variables, whose correct treatment depends on both the form of the model and the order of nesting \citep{hessdaly2026}.

The labels in common use do not make these distinctions. RU2, on its only reading that is a normalisation, identifies the utility-consistent form normalised at the top. RU1, by contrast, has been used for normalisation at the bottom, for the undivided form, and for a third form that also multiplies nest-level utilities by each nest's own parameter, sometimes within a single paper. NNNL is no better: the form it names is normalised, since it is identified, and what distinguishes it from the utility-consistent form is the omitted division rather than a missing normalisation. 
Its sources also differ in their treatment of nest-level utilities: \citet{daly1987} and NLOGIT's standard form include them unscaled, \citet{koppelman1998a} leave them out of their formulation, noting only that they may be included, and the RU1 of \citet{hensher2002}, which they identify with NNNL, multiplies them by each nest's own parameter, so that even this name does not identify a single model.

Reported results need the same care. Packages differ in whether they report scales or structural parameters, in whether structural parameters are measured against the root or against the parent, and hence in what range of values is consistent with utility maximisation, so that a value above one violates the conditions for consistency in one package and is the normal case in another. Estimates from different packages, or from different versions of the same package, should therefore be compared only after converting them to a common convention, and any published model should state the convention in which its structural parameters are reported.

\bibliographystyle{elsarticle-harv}
\bibliography{refs}

\end{document}